# Visual Mechanisms Inspired Efficient Transformers for Image and Video Quality Assessment

Junyong You[1], Zheng Zhang[2]

1. NORCE Norwegian Research Centre, Bergen, Norway
junyong.you@norceresearch.no

2. Hong Kong University of Science and Technology, Hong Kong, China
zzhanggq@connect.ust.hk

**Abstract.** Visual (image, video) quality assessments can be modelled by visual features in different domains, e.g., spatial, frequency, and temporal domains. Perceptual mechanism in the human visual system (HVS) play a crucial role in the generation of quality perception. This paper proposes a general framework for no-reference visual quality assessment using efficient windowed transformer architectures. A lightweight module for multi-stage channel attention is integrated into the Swin (shifted window) Transformer. Such module can represent appropriate perceptual mechanisms in image quality assessment (IQA) to build an accurate IQA model. Meanwhile, representative features for image quality perception in the spatial and frequency domains can also be derived from the IQA model, which are then fed into another windowed transformer architecture for video quality assessment (VQA). The VQA model efficiently reuses attention information across local windows to tackle the issue of expensive time and memory complexities of original transformer. Experimental results on both large-scale IQA and VQA databases demonstrate that the proposed quality assessment models outperform other state-of-the-art models by large margins. The complete source code will be published on Github.

## 1 Introduction

With the rapid development of mobile devices and social media platforms, e.g., TikTok, Instagram, we have seen an explosion of user-generated content (UGC) visual contents (images, videos). Evaluation of the perceived quality of UGC images/videos becomes a critical issue. Traditionally fully developed methods for visual quality assessment have been focused on full-reference or reduced-reference scenario, in which a distorted visual signal is assessed by fully or partially compared with the original undistorted signal [1]. However, no reference information for UGC image/video is available, as the distortions are often introduced by unprofessional producing methods, poor capturing environments and equipment, or other authentic artifacts. Thus, no-reference (NR) quality assessment naturally becomes the only choice for UGC quality assessment.

Early studies on NR image quality assessment (IQA) or video quality assessment (VQA) often targeted at certain distortion types, e.g., compression and transmission artefacts [2][3]. Later, in order to build a general-purpose quality assessment model, feature engineering is widely used [4][5]. Representative features relevant to quality perception are derived from visual signals in different domains, e.g., spatial, frequency and temporal domains. The features are then combined into quality prediction using either analytic approaches, or machine learning methods, or both. In [4], a widely referenced IQA model (BRISQUE) derives scene statistics of locally normalized luminance coefficients in the spatial domain, based on which support vector regressor (SVR) is used to predict an image quality score. Saad *et al.* [5] employed an analytic model (VBLIINDS) based on intrinsic statistical regularities mimicking distortions occurred on natural videos in VQA. To efficiently reduce complexity of processing long video sequences, Korhonen [6] proposed to model quality related video features on two levels: low complexity features on full video and high complexity features on representative frames only. The features are then fed into SVR for quality evaluation.

As deep learning is dominating computer vision tasks, it has also attracted research interests in image/video quality assessment. Earlier works on deep IQA models often follow the approach in image classification, e.g., adapting the CNN classification models for quality prediction. In [7], AlexNet pretrained on ImageNet is fine-tuned and used to extract image features for IQA. Hosu *et al.* [8] added several fully connected layers on the top of InceptionResNetV2 to develop an IQA model, which shows promising performance on their developed dataset KonIQ-10k. However, a potential issue with direct adaption of classification models for IQA lies in that image rescaling is required, as those classification models often accept fixed input sizes. Such image rescaling might not significantly affect the classification task, while it can have dramatic influence on perceived image quality. For example, viewers often prefer to watch high-dimension visual contents on large screens than low-resolution signals on small screens. Thus, image rescaling should be avoided in IQA models to prevent from introducing extra unnecessary distortions.

On the other hand, as a subjective concept, visual quality perception can be essentially determined by the intrinsic mechanisms in human visual system (HVS). For example, selective spatial attention can guide the quality assessment process. Viewers might pay different attention to different visual stimuli driven by the contents, viewing tasks and other factors in an overt or covert manner [9][10]. In addition, contrast sensitivity function (CSF) describes that the HVS presents different sensitivities to individual components in the visual stimuli [11][12]. It also affects quality perception as certain distortions do not trigger viewers' perception, e.g., a just noticeable distortion model is inspired by CSF [13]. As the transformer model can appropriately represent the selective attention mechanism, it has been applied in IQA. You *et al.* [14] proposed a hybrid model (TRIQ) using a transformer encoder based on features derived from a CNN backbone.

Furthermore, considering that both high-resolution and low-resolution features contain essential information for image quality perception, several models have been developed based on a multi-scale architecture. Ke *et al*. [15] proposed a multi-scale image quality transformer (MUSIQ) based on 3D spatial embedding and scale embedding at different granularities. In [16], a hierarchical module simulating spatial and channel attention is integrated into CNN backbone for multi-scale quality prediction. Wu *et al*. [17] proposed a cascaded architecture (CaHDC) to represent the hierarchical perception mechanism in HVS, and then pool the features extracted at different scales. In [18], a hyper network (hyperIQA) is built by aggregating multi-scale image features capturing both local and global distortions in a self-adaptive manner.

Compared to applying deep networks for IQA, VQA still does not widely benefit from the advances of deep learning models, due to the complexity and diversity of modelling spatial-temporal video characteristics. It is difficult to directly feed a long video sequence into deep networks consuming tremendous computing resources. For example, a rough estimation indicates that ResNet50-3D on 10s video (1024x768 @25FPS) requires ~14GB memory. Existing deep learning driven video models, e.g., for video classification, mainly perform frame down-scaling on short sequences to reduce the resource requirements [19]. Considering that image rescaling should be avoided in visual quality assessment, a feasible approach is to derive quality features from video frames first and then pool them in the temporal domain. This is actually a mainstream work in VQA [20]-[24]. For example, pretrained CNNs are often used to derive frame features and popular recurrent neural networks (RNN) or regressors are employed for temporal pooling, e.g., LSTM [20][21], GRU [22], 3D-CNN [23], and SVR [24]. However, those large-scale image datasets to pretrain CNNs were built for other purposes than quality assessment. Image features derived by the pretrained CNNs might not well represent quality related properties. Several studies intend to derive frame features by dedicated IQA models instead of pretrained CNNs, e.g., see [25][26]. Furthermore, as both spatial and temporal video characteristics contribute to video quality, quality features derived in the two domains have also been considered simultaneously in VQA models [27][28].

As a powerful presentation tool, transformer has already demonstrated outstanding performance in time-series data analysis, e.g., language modelling [29][30]. Naturally, it should also be considered in VQA as a temporal pooling approach. You *et al*. [25] has proposed a long short-term convolutional transformer for VQA. In [31], video frames are first preprocessed (spatial and temporal down-samplings) and then fed into a transformer with sequential time and space attention blocks.

Whereas, besides its excellent representation capability, transformer has an unignorable problem of quadratic time and memory complexity. Recently, several attempts have been made to develop efficient transformer models in computer vision and time-series data process. For example, the dot-product was replaced by locality-sensitivity hashing in Reformer to reduce the complexity [32]. Several efficient transformer models attempt to replace the full range attention modelling, as it significantly contributes to the high complexity of original Transformer. Zhu *et al*. [33] proposed a long-short Transformer by aggregating a long-range attention with dynamic projection for distant correlations and a short-term attention for fine-grained local correlations. Liu *et al*. [34] used a shifted windowing (Swin) scheme to decompose the full range attention into non-overlapping local windows while also connecting them crossing windows, reducing the complexity of transformer to linear.

In this work, we intend to propose a general, efficient framework for no-reference image and video quality assessment. Existing attention based IQA models mainly take spatial selective attention into account that can be easily modelled by transformer models. We propose a multi-stage channel attention model based upon Swin Transformer to predict image quality. Channel attention can appropriately simulate the contrast sensitivity mechanism. The multi-stage structure can, on one hand, produce more accurate image quality perception, as demonstrated in existing IQA models [15]-[17]. On the other hand, such multi-stage structure can also generate more representative features covering different resolution scales to provide solid foundation for VQA tasks. Furthermore, inspired by the kernel idea of CNNs and RNNs that the kernels (e.g., convolution kernels) can be reused crossing individual image patches or time-steps, we also propose a locally shared attention module followed by a global transformer for VQA. Our experiments on large-scale IQA and VQA datasets will demonstrate promising performance of the proposed models.

The remainder of the paper is organized as follows. Section 2 presents a multi-stage channel attention model using Swin Transformer as backbone for IQA. A VQA model using a locally shared attention module on frame quality features is detailed in Section 3. Experimental results and ablation studies are discussed in Section 4. Finally, Section 5 concludes the paper.

## 2 Multi-stage Channel Attention integrated Swin Transformer for IQA (MCAS-IQA)

In the proposed visual quality assessment framework, an IQA model needs to not only predict image quality accurately, but also serve as a general feature extractor for deriving quality features of video frames. Several mechanisms of the HVS play important roles in quality assessment, e.g., selective spatial attention, contrast sensitivity, adaptive scalable perception. Selective spatial attention is an inherent mechanism describing that viewers often allocate more attention to certain areas in the visual field than others. The attention allocation is determined by both viewing tasks and visual contents. Attention has been already widely studied in deep learning models, e.g., Swin Transformer and ViT [35] for image classification. However, the transformer based image models only consider spatial attention. In IQA scenario, another important mechanism, contrast sensitivity, plays a crucial role. Not all the distortions on an image can be perceived by the HVS. According to CSF, visual sensitivity is maximized at the fovea determined by the spatial frequency of visual stimuli and declined by eccentricity away from the gaze [12]. In other words, contrast change beyond certain frequency threshold will be imperceptible. Therefore, information in the frequency domain should also be considered in a deep learning driven IQA model.

Even though the frequency information is not explicitly represented by a deep learning model, it can be simulated by the channel outputs. For example, Sobel operation is a convolutional operator that can distinguish high frequency information (e.g., edges) from low frequency information (e.g., plain areas). Since contrast sensitivity can be explained as the visual stimuli with different frequency can trigger different attentional degrees in visual system. Thus, we assume that it can be simulated by channel attention.

Most deep learning models for image processing attempt to aggregate representative features in a manner of transferring spatial domain to channel. Consequently, the features for target tasks are often extracted from feature maps with high channel dimension while reduced spatial resolution. Such approach might be appropriate for classification or recognition purpose, while it can potentially lose information in the spatial domain in other scenarios. For example, feature fusion cross low-stage and high-stage feature maps is beneficial for object detection. We believe that spatial information also plays an important role in image quality perception, while crucial information in the spatial domain might be lost if only the highest stage feature map in a deep learning model is employed. Therefore, we attempt to use both low-stage and high-stage feature maps in the proposed IQA model.

Swin Transformer uses a windowed architecture for reducing the complexity. It also provides a hierarchical structure facilitating information aggregation over different stages. The proposed IQA model employs Swin Transformer as a backbone network. However, Swin Transformer only accepts fixed size of input image resolution. Due to the particularity of quality assessment as explained earlier, we intend to avoid image rescaling to unalter the perceived image quality. In order to adapt images with arbitrary resolutions to Swin Transformer, an adaptive spatial average pooling is performed during patch embedding. If we set the input size as 384×384 and patch size as 4×4 in the Swin Transformer backbone, the pooling kernel size and stride will be both set to 96(=384/4). After an image with arbitrary resolution has been divided into individual patches by a 2D convolution layer, the adaptative pooling layer will be performed to adapt the embedded patches to a fixed size for the next process. Considering that the 2D convolution with the kernel number being the embedding dimension has been performed first that can convert spatial image features to channel domain, we assume that such adaptive pooling does not introduce spatial information loss for quality perception.

Swin Transformer can decompose an input image into four stages, similarly to other CNNs. The feature maps at each stage have different spatial and channel resolutions. As explained earlier, we believe that attention over channels at each stage should both be considered, as the features over different channels can have different impacts on image quality perception. A lightweight channel attention block is built to represent the relative importance levels of different channel features. Thus, we propose to use a 1D dense layer (channel attention layer) with activation function of Sigmoid. A spatial average pooling is first performed on the feature maps at each stage, and then multiplied with the channel attention layer. In this way, higher attention weights will be assigned to more important information crossing channels during the training process.

Furthermore, we propose to share the channel attention layer across different stages for two advantages. First, sharing a dense layer can significantly reduce its weights. Second, we assume that the attentional behavior over channels is similar at different stages, then sharing the attention layer can force the model to learn such similarity. If we can assume that the features in different channels within a stage represent information from low frequency to high, and such pattern is similar in other stages. Subsequently, if the channel attention layer finds out that features in certain frequency bands are more important than others in one Swin stage, such behavior should hold in other stages. However, the feature maps have different channel numbers at individual stages. In order to share the channel attention layer, a bottleneck conv-layer is used on the feature maps in different stages. The filter number of the bottleneck conv-layer is set to the same as the channel number of the highest stage feature map, and the kernel size is 1×1. Subsequently, the bottleneck layer is performed on the feature maps at each stage to unify their channel dimension. The feature maps are then pooled by the global spatial pooling and finally multiplied with the shared channel attention layer. Assuming that the bottleneck layer has $K$ channels, a feature vector of size 1×$K$ will be produced at each stage.

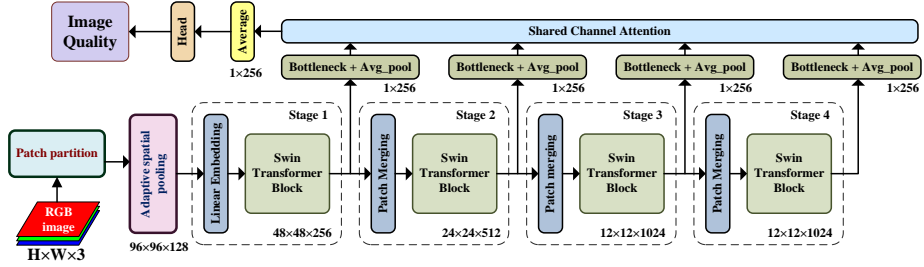

**Fig.1** Architecture of the MCAS-IQA model (Swin-B with input size of 384 used for Swin Transformer). The number below each node indicates the output shape.

Finally, a head layer is employed to derive image quality from the feature vectors. The head layer can be defined differently according to the prediction objective. In IQA tasks, MOS is often used representing the perceived quality by averaging over multiple subjective voters. Therefore, an IQA model can predict a single MOS value of an image directly, while it can also predict the distribution of quality scores over a rating scale, e.g., a commonly used five-point scale [bad, poor, fair, good, excellent]. In the former case, a dense layer with one unit and linear activation should be used as the head layer, and MSE between the ground-truth MOS values and predicted image quality is used as loss function. For the latter case, five units with Softmax activation are often used in the dense head layer, and the cross-entropy measuring distance on quality score distribution will be served as the loss function. In the experiments we will demonstrate the two cases on two IQA datasets.

There are two approaches to combine the feature vectors at different stages into image quality. One approach is to apply the head layer on the feature vector at each stage first and then average over stages, and the other approach is to average the feature vectors over stages first and then apply the head layer. Our experiments have shown that the first approach shows slightly better performance than the second approach. Therefore, we have employed the first approach in this work, and Fig. 1 illustrates the architecture of the proposed MCAS-IQA model.

As explained earlier, a feature vector with the size of 1×$K$ is derived at each stage, which conveys the most crucial and representative information for image quality perception in spatial and channel (frequency) domains by taking attention mechanism into account. Subsequently, by concatenating the feature vectors from the four stages, we obtain quality features over different scales and assume that the features represent general perceptual clues to be used in video quality assessment. In our experiments, the Swin-B method with input size of 384×384 is used as the Swin Transformer backbone. Consequently, $K$ equals to 256 and the concatenated quality feature vector on each frame has a shape of 1×1024.

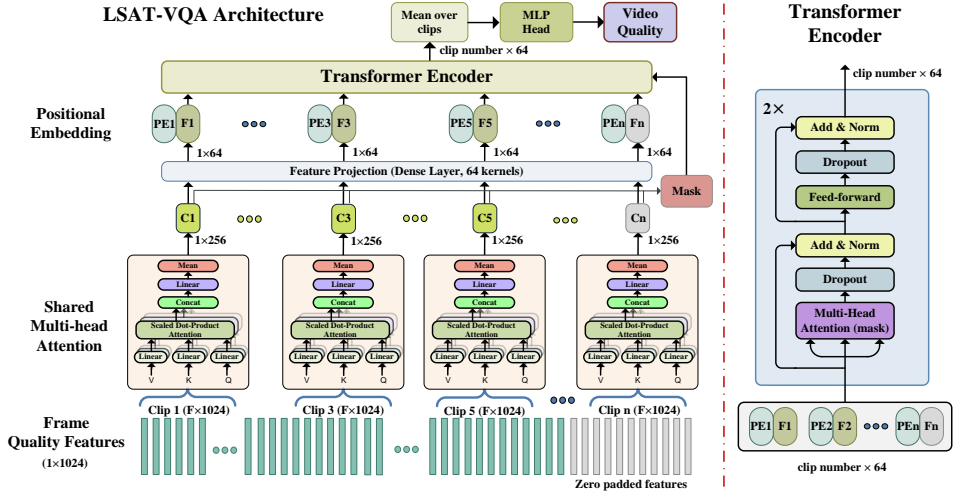

**Fig.2** Architecture of the LSAT-VQA model and transformer encoder. The number below each node indicates output shape.

## 3    Locally Shared Attention in Transformer for VQA (LSAT-VQA)

In the proposed visual quality assessment framework, the VQA model is to aggregate the quality features over all frames in a video sequence into a video quality score. In other words, video quality is derived by pooling the quality features in the temporal domain. Transformer has already demonstrated outstanding performance in processing time-series data, while its quadratic time and memory complexities also impedes a direct use of Transformer for quality prediction of long video sequences. However, the basic unit of generating video quality perception might be short clips, rather than individual frames. We have performed a simple questionnaire in an unreported subjective VQA experiment to collect viewers' opinion on whether they assessed video quality based on individual frames or short segments or the whole sequence. Most viewers have reported that they often retrospected to video clips, rather than frames, and combined the qualities of individual clips to determine the overall video quality. Thus, we propose to divide a video sequence into non-overlapping clips and process the quality features within individual clips, and finally the overall video quality is derived by pooling over clips. An overlapping division of clips has also been tested while no performance gain was obtained. Fig. 2 illustrates the architecture of the proposed LSAT-VQA model, and the details are presented as follows.

Attention mechanism plays an important role in VQA. It can be reasonably assumed that viewers pay varying attention to different temporal segments. For example, varying contents and quality levels can potentially attract unbalanced attention in time. Such uneven attention distribution can be modelled by the transformer encoder, as the self-attention module simulates the attention distribution over different segments in a video sequence. However, it is sophisticated to analyze the attentional behavior of quality assessment within a short segment. To our best knowledge, no psychovisual experiments have been conducted to study this issue. In this work, we hypothesize that attention mechanism can also be employed, i.e., different frames in a segment contribute unevenly to the quality at segment or clip level. More importantly, we presume that such attention modelling approach can be shared across different clips. In other words, an attention block will be reused in all the clips in a video sequence. Such idea has also been widely used in other deep learning models, e.g., the same kernel is used in CNN or RNN. The different aspect is that we intend to use only one single attention block, rather than multiple kernels as in CNN or RNN. This is to simplify the approach of modelling short-term attention within video segments, and it can also significantly reduce the complexity of the subsequent global transformer model.

The quality features from all the frames in a video clip are aggregated into a feature vector representing the perceptual information at clip level. Subsequently, the feature vectors from all the clips in a video sequence can be fed into a global transformer encoder to generate video quality prediction at sequence level.

The multi-head attention (MHA) layer proposed in the original Transformer model [29] is used to model the attention distribution over frames in a clip. The dimension of MHA is set to 256, and the head number is set to 4 in this study. Our ablation studies show that the two parameters only have minor impact on the performance of the proposed LSAT-VQA model. The MHA block produces a feature map with shape of $F \times 256$, where $F$ is the frame number in a clip. Subsequently, the average over frames is taken as the feature vector with a shape of $1 \times 256$ for this clip.

In addition, the situation that video sequences have different length should be handled in VQA models. Transformer can tackle varying input length. However, when performing batch training, padding operation is often used to pad video sequences with different numbers of frames to the same length in a batch. Zero padding is employed in this work. Consequently, the MHA block will also generate an all-zero feature vector on a clip with all zero padded frames. In order to exclude those padded frames from quality prediction, a masking operation has been performed by marking those all-zero feature vectors and assigning a very small value as the attention weights in the transformer encoder.

Subsequently, a global transformer encoder is applied to the feature vectors on all clips. Four key hyper-parameters are used to represent a transformer encoder, namely $L$ (layer), $D$ (model dimension), $H$ (head), and $d_{ff}$ (the feed-forward network). In our experiments, we have found that relatively shallow model parameters, e.g., {$L$=2, $D$=64, $H$=8, $d_{ff}$=256}, produce promising performance for VQA. We assume that this is due to the fact that the employed VQA datasets are often in small-scale. In order to match the dimension of the feature vectors (i.e., 256) derived from the shared MHA block to the model dimension $D$ of transformer encoder, we first perform a feature projection layer on the feature vectors. A dense layer with $D$ filters is used for feature projection.

As swapping the order of different clips can affect the perceived video quality, positional information should be retained in the transformer encoder. A learnable positional embedding is added to the projected feature vectors. The positional embedding layer is set sufficiently long to cover the maximal length of video clips in the used VQA databases, which can be truncated for shorter videos.

Eq. (1) roughly explains the shared MHA block ($S\_MHA$) on frame quality features ($QF$), the feature projection layer ($Proj$), and the positional embedding ($PE$), where $C$ denotes the number of non-overlapping clips in a video sequence.

$$\begin{cases} CF_j = S\_MHA(QF_j), & QF_j \in R^{1 \times K} \\ F_j = Proj(CF_j), & CF_j \in R^{1 \times 256} \\ Z_0 = [F_1 + PE_1; \cdots F_C + PE_C], & F_j \in R^{1 \times D}, \; PE \in R^{C \times D} \end{cases} \quad (1)$$

It is noted that no extra classification token has been added to the beginning of the project features, which is different from traditional transformer models for classification, e.g., BERT [30] and ViT [35]. Instead, the average of transformer encoder output along the attended dimension will be fed into the next process.

Each encoder layer in transformer contains $L$ sublayers for multi-head attention ($MHA$) with mask and a position-wise feed-forward layer ($FF$). Layer normalization ($LN$) is performed on the residual connection around the sublayers. Is should be noted that the $MHA$ in the transformer encoder is different from the earlier shared MHA block. Eq. (2) explains the pseudo approach of transformer encoder based on the output from the shared MHA block.

$$\begin{cases} Z'_l = LN(MHA(Z_{l-1}) + Z_{l-1}) \\ Z_l = LN(FF(Z'_l) + Z'_l) \end{cases} \quad l = 1, \cdots L \quad (2)$$

The encoder outputs at the last encoder layer are then averaged over all the clips to produce a quality vector with size of $1 \times D$. Such quality vector is supposed to contain the most representative information for quality perception at the sequence level.

Finally, a multi-layer perceptron head layer is performed on the quality vector to predict video quality. The head layer consists of two dense layers and a dropout layer in between. Following other transformer architectures, e.g., BERT, ViT, Swin Transformer, GELU activation is used in the first dense layer with $d_{ff}$ units. As the goal of VQA in this task is to predict a single quality score, the second dense layer uses only one unit and linear activation. Accordingly, MSE is chosen as the loss function to measure the distance between the ground-truth MOS values and predicted video quality.

## 4  Experiments and discussions

**4.1 IQA and VQA datasets**

Appropriate datasets are crucial for training deep learning models. Several large-scale databases with authentic distortions for IQA [8][36] and VQA [37][38] have been used in this work.

To the best of our knowledge, there are currently three large-scale IQA datasets, including SPAQ [36], KonIQ-10k [8] and LIVE-FB [39]. SPAQ containing 11,125 images with diverse resolutions was produced in a controlled laboratory environment. Hosu *et al*. conducted two large-scale quality assessment experiments on crowdsourcing platforms for IQA and VQA, respectively. Two datasets were published, namely the KonIQ-10k IQA dataset [8] containing over 10,000 images with a constant resolution of 1024×768 and KonViD-1k VQA dataset [37] consisting of 1,200 videos with varying resolutions and durations. It should be noted that the quality score distributions from the participants have also been published in the KonIQ-10k dataset. Therefore, we can use the IQA model to predict the score distribution by using five units and Softmax activation in the head layer. On the other hand, only MOS values were published in the SPAQ dataset, and a single unit with linear activation is used in the head layer. In addition, although LIVE-FB [39] contains 39,806 images, the largest (by far) IQA dataset, it is actually assembled from several existing databases purposed for other tasks than IQA. Consequently, the MOS values of 92% images in LIVE-FB locate in a narrow range of [60, 80] out of the full range [0, 100], which is inappropriate for training IQA models due to lack of representative distortions of diverse image quality levels. Thus, the evaluation of IQA models has been mainly conducted on KonIQ-10k and SPAQ in our experiments. In addition, another large-scale VQA dataset, YouTube-UGC [38], consists of 1,288 videos with authentic distortions. The evaluation of VQA models has been mainly performed on KonViD-1k and YouTube-UGC.

Differently from other computer vision tasks where individual databases can be combined directly, e.g., image classification and object detection, quality assessment datasets might be incompatible from each other. A tricky issue is quality level calibration between separate datasets. For example, two images with similar quality level might be assigned with dramatically different quality values in two subjective experiments, because the calibration levels deviate significantly. Some researchers have attempted to combine multiple datasets to train a single IQA model, e.g., UNIQUE model [40] and MDTVSFA [41]. While in this work, we still concentrate on training IQA/VQA models on individual datasets.

Subsequently, the individual datasets were randomly split into train, validation and test sets following the standard protocol. However, in order to avoid the long tail issue, we first roughly divided the samples in each dataset into two complexity categories (low and high) based on the spatial perceptual information (SI) and temporal information (TI) defined by the ITU Recommendation [42]. The samples in each category were then roughly divided into five quality subcategories based on their MOS values. Finally, we randomly chose 80% of the samples in each subcategory as train set, 10% as validation set and the rest 10% as test set. Such random split has been repeated for ten times.

Furthermore, three other small-scale quality assessment datasets with authentic distortions, namely CID2013 [43] and CLIVE [44] for IQA, and LIVE-VQC [45] for VQA, have also been included to evaluate the generalization capabilities of different models. It should be noted that individual datasets use different ranges of MOS values, e.g., [0, 100] in SPAQ, CLIVE and LIVE-VQC, and [1, 5] in CID2013, KonIQ-10k, KonViD-1k and YouTube-UGC. For convenient evaluation across datasets, all the MOS values were linearly normalized into the range of [1, 5] in our experiments.

Table 1. Evaluation results (average and standard deviation) on the test set on individual IQA datasets

| Models | KonIQ-10k | | | SPAQ | | |
|---|---|---|---|---|---|---|
| | PLCC↑ | SROCC↑ | RMSE↓ | PLCC↑ | SROCC↑ | RMSE↓ |
| DeepBIQ | 0.873 ±0.021 | 0.864 ±0.037 | 0.284 ±0.029 | 0.858 ±0.027 | 0.861 ±0.028 | 0.389 ±0.035 |
| Koncept512 | 0.916 ±0.116 | 0.909 ±0.085 | 0.267 ±0.094 | 0.831 ±0.097 | 0.830 ±0.080 | 0.384 ±0.060 |
| TRIQ | 0.922 ±0.018 | 0.910 ±0.011 | 0.223 ±0.030 | 0.916 ±0.027 | 0.925 ±0.015 | 0.324 ±0.021 |
| MUSIQ | 0.925 ±0.011 | 0.913 ±0.029 | 0.216 ±0.026 | 0.920 ±0.010 | 0.918 ±0.006 | 0.339 ±0.030 |
| AIHIQnet | 0.929 ±0.020 | 0.915 ±0.014 | 0.209 ±0.022 | 0.929 ±0.022 | 0.925 ±0.019 | 0.326 ±0.027 |
| CaHDC | 0.856 ±0.027 | 0.817 ±0.025 | 0.370 ±0.041 | 0.824 ±0.030 | 0.815 ±0.019 | 0.486 ±0.068 |
| hyperIQA | 0.916 ±0.030 | 0.907 ±0.027 | 0.242 ±0.012 | 0.910 ±0.026 | 0.915 ±0.020 | 0.329 ±0.028 |
| MCAS-IQA | *0.956* ±0.020 | *0.944* ±0.015 | *0.163* ±0.023 | *0.933* ±0.021 | *0.926* ±0.021 | *0.304* ±0.020 |

### 4.2 Training approach

The performance of deep learning models is heavily dependent on training approach. For training the MCAS-IQA model, transfer learning for the Swin Transformer backbone pretrained on ImageNet was applied. However, there are no available large-scale pretrained models for video processing. Thus, the VQA models have been trained from scratch.

In our experiments, the training process of each model was performed in two phases: pretrain and finetune. In both two phases, a learning rate scheduler of cosine decay with warmup was used in the Adam optimizer. The base learning rate was determined from a short training by changing learning rates and monitoring the model results. We have found that 5e-5 and 1e-6 produced the best performance in the pretrain and finetune phases in training the MCAS-IQA model. For training the LSAT-VQA model, we have finally used 1e-3 and 5e-5 as base learning rates in pretrain and finetune, respectively, together with a large batch size.

Data augmentation is widely used in model training to increase the dataset size. However, due to the particularity of quality assessment, most image or video augmentation methods might be inappropriate as they can affect the perceived quality. We have investigated popular image augmentations, e.g., transformation, adding noise as those implemented in [46], and found that only horizontal flip has no significant impact on image quality. For LSAT-VQA training, the quality features on horizontally flipped frames were also computed and included the training process. In addition, 25% randomly chosen videos in the train set were reversed in each batch in the training, i.e., the quality features ordered from the last frame to the first are fed into LSAT-VQA. Even though such approach seems to be conflict with the conclusion that swapping video clips can affect video quality, we have found that using reversed video in training improves the performance slightly.

Table 2. Ablation studies on MCAS-IQA

| Ablations | PLCC | SROCC | RMSE |
|---|---|---|---|
| *Original MCAS-IQA* | *0.945* | *0.935* | *0.234* |
| 1) No multi-stage, no channel attention | 0.913 | 0.903 | 0.302 |
| 2) No multi-stage, with channel attention | 0.936 | 0.919 | 0.280 |
| 3) Multi-stage, without channel attention | 0.925 | 0.927 | 0.258 |

Table 3. Model generalization studies

| Datasets | PLCC | SROCC | RMSE |
|---|---|---|---|
| KonIQ-10k vs. SPAQ | 0.861 | 0.875 | 0.454 |
| SPAQ vs. KonIQ-10k | 0.834 | 0.829 | 0.309 |
| Tested on CID2013 | 0.858 | 0.853 | 0.602 |
| Tested on CLIVE | 0.864 | 0.877 | 0.584 |

During the training process, the Pearson correlation (PLCC) between the predicted quality scores and ground-truth MOS on the validation set was used to monitor the performance of trained models. Early stop has also been employed to avoid overfitting. All the models were trained on the individually split train sets on two GeForce RTX 3090 GPUs and the best weights were determined by PLCC values on the validation sets. Subsequently, the model performance was evaluated on the test sets in terms of three criteria: PLCC, Spearman rank-order correlation (SROCC) and root mean squared error (RMSE) between the predicted quality scores and the ground-truth MOS values.

**4.3 Comparison and ablations of IQA models**

Table 1 reports the evaluation results of the proposed MCAS-IQA model compared against other state-of-the-art models, including several deep learning driven models: DeepBIQ [7], Koncept512 [8], TRIQ [14], MUSIQ [15], AIHIQnet [16], CaHDC [17] and hyperIQA [18]. Other detailed results can be found on the code repository page. The comparison results demonstrate that MCAS-IQA significantly outperforms other compared models. We assume that the reasons are two twofold. First, Swin Transformer as the backbone with large-scale pretrain has outstanding representative capability. Second, we presume that the proposed multi-stage architecture of shared channel attention can appropriately simulate the perceptual mechanisms of IQA.

In order to fully reveal the benefit of using multi-stage channel attention architecture, we have conducted three ablation studies: 1) No multi-stage and no-channel attention: the output from the last Swin Transformer block is averaged and then fed into the head layer directly; 2) No multi-stage but with channel attention: the channel attention layer is performed on the output from the last Swin Transformer block; and 3) Multi-stage without channel attention: the outputs from the four Swin Transformer blocks are averaged and then fed into the head layer. Table 2 presents the results in terms of the averaged criteria between KonIQ-10k and SPAQ. The ablation studies confirm that using multi-stage and shared channel attention architecture dramatically improve the performance on IQA. Furthermore, it seems that using channel attention offers more benefit than the multi-stage approach in the proposed MCAS-IQA architecture.

Subsequently, we have conducted another experiment to evaluate the generalization capability of MCAS-IQA, as reported in Table 3. The model was first trained on one dataset and then tested on another dataset, e.g., KonIQ-10k vs. SPAQ indicates MCAS-IQA was trained on KonIQ-10k and tested on entire SPAQ, vice versa. We have also trained MCAS-IQA on the combination of KonIQ-10k and SPAQ and then tested on entire CID2013 and CLIVE datasets, respectively. According to the evaluation results, MCAS-IQA shows strong generalization capability across datasets and subjective experiments, which provides a solid foundation to use MCAS-IQA as a general model to derive quality features for VQA. Thus, MCAS-IQA trained on the combination of KonIQ-10k and SPAQ has been employed to compute the quality features of every frame in video sequences for the VQA experiments.

### 4.3 Experiments on VQA models

For VQA, LSAT-VQA has been compared against the following state-of-the-art models: TLVQM [6], VSFA [22], 3D-CNN [23], VIDEAL [24], LSCT [25], ST-

**Table 4 (a).** Evaluation results (average and standard deviation) on the test set on individual VQA datasets

| Models | KonIQ-10k | | | SPAQ | | |
|---|---|---|---|---|---|---|
| | PLCC↑ | SROCC↑ | RMSE↓ | PLCC↑ | SROCC↑ | RMSE↓ |
| TLVQM | 0.76±0.02 | 0.76±0.02 | 0.42±0.03 | 0.68±0.03 | 0.65±0.03 | 0.49±0.02 |
| VSFA | 0.80±0.03 | 0.80±0.02 | 0.40±0.03 | 0.77±0.03 | 0.79±0.04 | 0.41±0.03 |
| 3D-CNN | 0.80±0.02 | 0.81±0.03 | 0.38±0.02 | 0.71±0.03 | 0.72±0.04 | 0.44±0.02 |
| VIDEAL | 0.67±0.02 | 0.65±0.02 | 0.50±0.03 | 0.66±0.03 | 0.68±0.03 | 0.49±0.02 |
| LSCT | 0.84±0.02 | ***0.85±0.02*** | 0.34±0.03 | 0.82±0.04 | 0.82±0.03 | 0.39±0.02 |
| ST-3DDCT | 0.73±0.02 | 0.74±0.02 | 0.44±0.03 | 0.48±0.04 | 0.49±0.04 | 0.59±0.03 |
| StarVQA | 0.80±0.04 | 0.80±0.03 | 0.40±0.02 | 0.80±0.06 | 0.78±0.03 | 0.44±0.02 |
| LSAT-VQA | ***0.85±0.02*** | 0.84±0.02 | ***0.33±0.02*** | ***0.85±0.03*** | ***0.83±0.03*** | ***0.38±0.03*** |

**Table 4 (b).** Evaluation results (average and standard deviation) on the combined dataset

| Models | Combined test set | | |
|---|---|---|---|
| | PLCC↑ | SROCC↑ | RMSE↓ |
| TLVQM | 0.71±0.02 | 0.74±0.02 | 0.46±0.02 |
| VSFA | 0.79±0.02 | 0.79±0.03 | 0.44±0.03 |
| 3D-CNN | 0.71±0.02 | 0.72±0.02 | 0.49±0.02 |
| VIDEAL | 0.67±0.02 | 0.67±0.12 | 0.57±0.02 |
| LSCT | 0.81±0.04 | 0.80±0.03 | 0.41±0.05 |
| ST-3DDCT | 0.59±0.03 | 0.60±0.02 | 0.58±0.02 |
| StarVQA | 0.77±0.05 | 0.79±0.04 | 0.48±0.04 |
| LSAT-VQA | ***0.84±0.02*** | ***0.83±0.04*** | ***0.40±0.04*** |

3DDCT [27] and StarVQA [31]. Table 4 reports the average and standard deviation of the evaluation criterions on individual VQA datasets and their combination. When using the combined dataset, the train sets, validation sets from KonViD-1k and YouTube-UGC were combined respectively in each of the ten random splits to train the VQA models, and then they were evaluated on the combined test sets. According to the comparison results, the proposed LSAT-VQA model show dramatically better performance than the compared models by a large margin. We assume that the reasons are still twofold. First, the proposed MCAS-IQA model that has been dedicatedly trained for IQA provide a solid foundation to derive frame features representing perceptual quality information. Next, the locally shared attention module based on transformer can appropriately represent the VQA mechanism in temporal domain, especially the clipping operation and using transformer for temporal polling over clips are in accordant with the subjective process performed by human viewers in VQA.

Table 5 also reports the evaluation result of the models trained on the combined KonViD-1k and YouTube-UGC dataset and then tested independently on the LIVE-VQC dataset. Due to high diversity of video contents and quality levels, it is expected that VQA models should have weak generalization capability. However, the proposed LSAT-VQA model still shows promising potential to become a general VQA model that can be applied to a wide range of video contents, resolutions, and distortions.

In addition, we have also compared the inference time of different VQA models on LIVE-VQC dataset. Average inference time including feature derivation and quality prediction was computed. The proposed VQA model is still the fastest, e.g., it takes ~300ms for a HD video with 5s duration. Table 5 also reports a rough comparison result on inference time of other models related to the proposed LSAT-VQA model. Furthermore, the LSAT-VQA is lightweight, and the size of the model file is only ~5MB.

Considering that the complexity of the global transformer is directly determined by the input length, sharing and reusing the local MHA in all the non-overlapping clips can dramatically reduce the length of the transformer input. Thus, it is important to dig into the architecture of building transformer on top of locally shared MHA. The first three ablation experiments are focused on comparing variations of the shared MHA layer, its influence on VQA, and clip length. In addition, another ablation has also been conducted to study the hyper-parameter settings of the global transformer. Finally, in order to evaluate the influence of dedicated IQA models on video quality perception against other pretrained image models, the ImageNet pretrained ResNet50 is used to extract features from video frames and then fed into the proposed LSAT-VQA model in the fifth ablation. All the ablation studies were performed on the combined dataset of KonViD-1k and YouTube-UGC. The following summarizes the ablation experiments and Table 6 reports the average results over the ten random splits.

**Table 5.** Performance evaluation and inference time relative to LSAT-VQA on LIVE-VQC dataset

| Models | PLCC | SROCC | RMSE | Time |
|---|---|---|---|---|
| TLVQM | 0.432 | 0.450 | 0.593 | 4.8 |
| VSFA | 0.663 | 0.640 | 0.540 | 2.3 |
| 3D-CNN | 0.598 | 0.603 | 0.522 | 2.7 |
| VIDEAL | 0.495 | 0.445 | 0.601 | 7.2 |
| LSCT | 0.702 | 0.730 | 0.497 | 3.4 |
| ST-3DDCT | 0.474 | 0.489 | 0.607 | 1.4 |
| StarVQA | 0.667 | 0.705 | 0.528 | 1.5 |
| LSAT-VQA | *0.724* | *0.726* | *0.490* | *1.0* |

1) Set the dimension and head number in the locally shared MHA layer to [64, 4] and [1024, 8], respectively;

2) Do not use the locally shared MHA layer, i.e., the frame quality features are fed into the transformer encoder directly;

3) Set the clip length to 8 and 128, respectively;

4) Test two hyper-parameter settings in the global transformer;

5) Feed the ResNet50 features into LSAT-VQA.

According to the ablation studies, the setting of the locally shared MHS is not very crucial for the performance of LSAT-VQA. However, a sole transformer encoder without the locally shared MHA significantly drops the performance. This partly confirms our assumption that the proposed local-global structure is appropriate for VQA, which is also in line with the observation that viewers often gauge video quality based on temporal pooling over clips, rather than frames. Furthermore, it is observed that relatively short clip length seems to produce slightly better performance in the proposed LSAT architecture. Considering that a longer clip length can reduce more complexity of transformer, it is worth finding an optimal balance between clip length and model accuracy. In our experiments, we have found that clip length 32 generally offers the optimal performance in video quality prediction.

In addition, even though large transformer models often show outstanding performance in language modelling and computer vision tasks, this is not the case in our experiments. For example, the deep transformer encoder in ablation 4) does not produce higher performance than the shallow one. Our hypothesis is that training on relatively small-scale VQA datasets cannot take the full advantage of large models. Finally, the last ablation study demonstrates that a dedicated and retrained IQA model definitely benefit VQA tasks more than a pretrained image classification model.

**Table 6.** Evaluation results of LSAT in the ablation studies

|   | Ablations | PLCC | SROCC | RMSE |
|---|---|---|---|---|
|   | ***Original LSAT-VQA*** | 0.843 | 0.834 | ***0.398*** |
| 1) | *MHA_S*: [64, 4] | 0.841 | 0.834 | 0.401 |
|   | *MHA_S*: [1024, 8] | 0.836 | 0.830 | 0.412 |
| 2) | No locally shared attention | 0.765 | 0.796 | 0.451 |
| 3) | Clip length: 8 | ***0.849*** | 0.834 | 0.399 |
|   | Clip length: 128 | 0.838 | 0.831 | 0.404 |
| 4) | Transformer: [2, 32, 4, 64] | 0.842 | ***0.835*** | 0.399 |
|   | Transformer: [4, 256, 8, 1024] | 0.836 | 0.829 | 0.308 |
| 5) | LSAT on ResNet50 features | 0.829 | 0.827 | 0.426 |

## 5      Conclusion

This paper proposed an efficient and general framework for no-reference image and video quality assessment by windowed transformers. Considering several important visual mechanisms in IQA, a multi-stage channel attention model MCAS-IQA built on Swin Transformer as backbone has been developed. MCAS-IQA demonstrates outstanding performance to predict perceived image quality whilst providing generic image quality features for VQA. Subsequently, inspired by the subjective mechanism that video quality is potentially a fusion of quality assessment over short video segments, we proposed a locally shared attention module followed by global transformer for VQA. By reusing local attention crossing video clips, the proposed LSAT-VQA model accurately predict video quality in an efficient manner. Complete comparison experiments have demonstrated outstanding performance of the proposed IQA and VQA models. The ablation studies have also revealed influence of individual building components driven by relevant visual mechanisms on image and video quality assessment.

## References

1. Wang Z., Sheikh H. R., and Bovik A. C.: Objective video quality assessment. in *The Handbook of Video Databases: Design and Applications*. CRC Press, 1041-1078 (2003).
2. Sazaad P.Z.M., Kawayoke Y., and Horita Y.: No reference image quality assessment for JPEG2000 based on spatial features. *Signal Process. Image Commun*. 23(4): 257-268 (2008).
3. Shahid M., Rosshold A., Lövström B., and Zepernick H.J.: No-reference image and video quality assessment: a classification and review of recent approaches. *EURASIP J. Image Video Proces*. 40:,1-40 (2014).
4. Mittal A., Moorthy A.K., and Bovik A.C.: No-reference image quality assessment in the spatial domain. *IEEE Trans. Image Process*., 21(12), 4695-4708 (2012).
5. Saad M. A., Bovik A. C., and Charrier C.: Blind prediction of natural video quality. *IEEE Trans. Image Process*., 23(3), 1352-1365 (2014).
6. Korhonen J.: Two-level approach for no-reference consumer video quality assessment. *IEEE Trans. Image Process*., 28(12), 5923-5938 (2019).
7. Bianco S., Celona L., Napoletano P., and Schettini R.: On the use of deep learning for blind image quality assessment. *Signal, Image and Video Process*., 12, 355-362 (2018).
8. Hosu V., Lin H., Sziranyi T., and Saupe D.: KonIQ-10k: An ecologically valid database for deep learning of blind image quality assessment. *IEEE Trans. Image Process.*, 29, 4041-4056, (2020).
9. Itti L., and Koch C. 2001. Computational modelling of visual attention. *Nat. Rev. Neurosci*., 2, 194-203.
10. Engelke U., Kaprykowsky H., Zepernick H.-J., and Ndjiki-Nya P.: Visual attention in quality assessment. *IEEE Signal Process. Mag.*, 28(6), 50-59 (2011).
11. Kelly H.: Visual contrast sensitivity, *Optica Acta: Int. J. Opt.*, 24(2), 107-12 (1977).
12. Geisler W. S., and Perry J. S.: A real-time foveated multi-resolution system for low-bandwidth video communication. *SPIE Human Vision Electron. Imaging*, 3299, 294-305, San Jose, CA, USA, (1998).

13. Zhang X., Lin W., and Xue P.: Just-noticeable difference estimation with pixels in images. *J. Vis. Commun.*. 19(1), 30-41 (2007).
14. You J., and Korhonen J.: Transformer for image quality assessment. *IEEE. Int. Conf. Image Process. (ICIP)*, Anchorage, Alaska, USA (2021).
15. Ke J., Wang O., Wang Y., Milanfar P., and Yang F.: MUSIQ: Multi-scale image quality Transformer. *IEEE/CVF Int. Conf. Comput. Vis.. (ICCV)*, Virtual (2021).
16. You J., and Korhonen J.: Attention integrated hierarchical networks for no-reference image quality assessment. *J. Vis. Commun.*, 82 (2022).
17. Wu J., Ma J., Liang F., Dong W., Shi G., and Lin W.: End-to-end blind image quality prediction with cascaded deep neural network. *IEEE Trans. Image Process.*, 29: 7414-7426 (2020).
18. Su S., Yan Q., Zhu Y., Zhang C., Ge X., Sun J., and Zhang Y.: Blindly assess image quality in the wild guided by a self-adaptive hyper network. *IEEE Comput. Soc. Conf. Comput. Vis. Pattern Recognit. (CVPR)*, Virtual (2020).
19. Karpathy A., Toderici G., Shetty S., Leung T., Sukthankar R., and Li F.-F.: Large-scale video classification with convolutional neural networks. *IEEE Comput. Soc. Conf. Comput. Vis. Pattern Recognit. (CVPR)*, Columbus, OH. USA (2014).
20. Varga D. and Szirányi T.: No-reference video quality assessment via pretrained CNN and LSTM networks. *Signal Image Video P.*, 13, 1569–1576 (2019).
21. Korhonen J., Su Y., and You J.: Blind natural video quality prediction via statistical temporal features and deep spatial features. *ACM Int. Conf. Multimed. (MM)*, Seattle, United States (2020).
22. Li D., Jiang T., and Jiang M.: Quality assessment of in-the-wild videos. *ACM Int. Conf. Multimed. (MM)*, Nice France (2019).
23. You J. and Korhonen J.: Deep neural networks for no-reference video quality assessment. *IEEE. Int. Conf. Image Process. (ICIP)*, Taipei, Taiwan (2019).
24. Tu Z., Wang Y., Birkbeck N., Adsumilli B., and Bovik A.C.: UGC-VQA: Benchmarking blind video quality assessment for user generated content. *IEEE Trans. Image Process.*, 30, 4449-4464 (2021).
25. You J.: Long short-term convolutional transformer for no-reference video quality assessment. *ACM Int. Conf. Multimed. (MM)*, Chengdu, China (2021).
26. Göring S., Skowronek J., and Raake A.: DeViQ - A deep no reference video quality model. *Proc. Human Vision and Electronic Imaging (HVEI),* Burlingame, California USA (2018).
27. Li X., Guo Q., and Lu X.: Spatiotemporal statistics for video quality assessment. *IEEE Trans. Image Process.*, 25(7), 3329-3342 (2018).
28. Lu Y., Wu J., Li L., Dong W., Zhang J., and Shi G.: Spatiotemporal representation learning for blind video quality assessment. *IEEE Trans. Circuits Syst. Video Technol..* 32, (6), 3500 - 3513 (2021)
29. Vaswani A., Shazeer N., Parmar N., Uszkoreit J., Jones L., Gomez A.N., Kaiser L., and Polosukhin I.: Attention is all your need. *Adv. Neural Inf. Process. Syst. (NIPS)*, Long Beach, CA, USA (2017).
30. Devlin J., Chang M.-W., Lee K., and Toutanova K.: BERT: Pre-training of deep bidirectional transformers for language understanding. *Proc. NAACL-HLT*, 1:4171–4186, Minneapolis, Minnesota, USA (2019).
31. Xing F., Wang Y-G., Wang H., Li L., and Zhu G.: StarVQA: Space-time attention for video quality assessment. https://doi.org/10.48550/arXiv.2108.09635 (2021).
32. Kitaev N., Kaiser L., and Levskaya A.: Reformer: The efficient Transformer. *Int. Conf. Learn. Represent. (ICLR)*, Virtual (2020).

33. Zhu C., Ping W., Xiao C., Shoeybi M., Goldstein T., Anandkumar A., and Catanzaro B.: Long-short Transformer: Efficient Transformers for language and vision. *Adv. Neural Inf. Process. Syst. (NeurIPS)*, Virtual (2021).
34. Liu Z., Lin Y., Cao Y., Hu H., Wei Y., Zhang Z., Lin S., and Guo B.: Swin Transformer: Hierarchical vision Transformer using shifted windows. *IEEE/CVF Int. Conf. Comput. Vis.. (ICCV)*, Virtual (2021).
35. Dosovitskiy A., Beyer L., Kolesnikov A., Weissenborn D., Zhai X., Unterthiner T., Dehghani M., Minderer M., Heigold G., Gelly S., Uszkoreit J., and Houlsby N.: An image is worth 16x16 words: Transformers for image recognition at scale. *Int. Conf. Learn. Represent. (ICLR)*, Virtual (2021).
36. Fang Y., Zhu H., Zeng Y., Ma K., and Wang Z.: Perceptual quality assessment of smartphone photography. *IEEE Comput. Soc. Conf. Comput. Vis. Pattern Recognit. (CVPR)*, Virtual (2020).
37. Hosu V., Hahn F., Jenadeleh M., Lin H., Men H., Sziranyi T., Li S., and Saupe D.: The Konstanz natural video database (KoNViD-1k). *Int. Conf. on Quality of Multimedia Experience (QoMEX)*, Erfurt, Germany (2017).
38. Wang Y., Inguva S., and Adsumilli B.: YouTube UGC dataset for video compression research. *Int. Workshop Multimed. Signal Process. (MMSP)*, Kuala Lumpur, Malaysia (2019).
39. Ying Z., Niu H., Gupta P., Mahajan D., Ghadiyaram D., and Bovik A.C.: From patches to pictures (PaQ-2-PiQ): Mapping the perceptual space of picture quality. *IEEE Comput. Soc. Conf. Comput. Vis. Pattern Recognit. (CVPR)*, Virtual (2020).
40. Zhang W., Ma K., Zhai G., and Yang X.: Uncertainty-aware blind image quality assessment in the laboratory and wild. *IEEE Trans. Image Process*, 30, 3474-3486 (2021).
41. Li D., Jiang T., and Jiang M.: Unified quality assessment of in-the-wild videos with mixed datasets training. *Int. J. Comput. Vis.*, 129, 1238–1257 (2021).
42. ITU-T Recommendation P.910. Subjective video quality assessment methods for multimedia applications," ITU (2008).
43. Virtanen T., Nuutinen M., Vaahteranoksa M., Oittinen P., and Häkkinen J.: CID2013: A database for eval-uating no-reference image quality assessment algorithms. *IEEE Trans. Image Process*, 24(1), 390-402 (2015).
44. Ghadiyaram D., and Bovik A.C.: Massive online crowdsourced study of subjective and objective picture quality. *IEEE Trans. Image Process*, 25(1), 372-387 (2016).
45. Sinno Z. and Bovik A.C.: Large-scale study of perceptual video quality. *IEEE Trans. Image Process.*, 28(2), 612-627 (2019).
46. A.B. Jung, K. Wada, J. Crall, *et al*., Imgaug, https://github.com/aleju/imgaug.